\documentclass[pdftex,twocolumn,epjc3]{svjour3}          % twocolumn
\RequirePackage[T1]{fontenc}

\smartqed  % flush right qed marks, e.g. at end of proof

\RequirePackage{overpic}
\RequirePackage{graphicx}
\RequirePackage{mathptmx}      % use Times fonts if available on your TeX system
\RequirePackage{flushend}
\RequirePackage[numbers,sort&compress]{natbib}
\RequirePackage[colorlinks,citecolor=blue,urlcolor=blue,linkcolor=blue]{hyperref}

\usepackage{amsmath}
\usepackage{amssymb}
\usepackage{amsfonts}

\journalname{Eur. Phys. J. C}

\begin{document}

\title{Monte Carlo study of particle production in diffractive proton-proton collisions at $\sqrt s$ = 13\;TeV with the very forward detector combined with central information}

\author{Qi-Dong Zhou\thanksref{e1,addr1}
        \and
        Yoshitaka Itow\thanksref{addr1,addr2}
        	\and
	Hiroaki Menjo\thanksref{addr3}
	\and
	Takashi Sako\thanksref{addr1,addr2}
}

\thankstext{e1}{e-mail: zhouqidong@isee.nagoya-u.ac.jp}

\institute{Institute for Space-Earth Environmental Research, Nagoya University, Nagoya, Japan\label{addr1}
          \and
          Kobayashi-Maskawa Institute, Nagoya University, Nagoya, Japan\label{addr2}
          \and
          Graduate School of Science, Nagoya University, Nagoya, Japan\label{addr3}
}

\date{Received: date / Accepted: date}

\maketitle
%=========================================================================================
%=========================================================================================
\begin{abstract}
Very forward (VF) detectors in hadron colliders, having unique sensitivity to diffractive processes, can be a powerful 
tool for studying diffractive dissociation by combining them with central detectors. 
Several Monte Carlo simulation samples in $p$--$p$ collisions at $\sqrt s = 13$ TeV were analyzed, and 
different nondiffractive and diffractive contributions were clarified through differential cross sections of
forward neutral particles. 
Diffraction selection criteria in the VF-triggered-event samples were determined by using the central track information.
The corresponding selection applicable in real experiments has $\approx$100\% purity and 30\%--70\% efficiency. 
Consequently, the central information enables classification of the forward productions into diffraction and nondiffraction categories; in particular, most of the surviving events from the selection belong to low-mass diffraction events at $\log_{10}(\xi_{x}) < -5.5$. 
Therefore, the combined method can uniquely access the low-mass diffraction regime experimentally.
  
\end{abstract}
%=========================================================================================
%=========================================================================================
\section{Introduction}
Inelastic hadronic collisions are usually classified into {\em soft processes\/} and {\em hard processes\/}.
Most of the hard processes can be treated within a theoretical framework based on perturbative quantum chromodynamics (QCD) owing to the large momentum transfer $t$. 
However, perturbative QCD is inadequate for describing soft processes such as diffractive dissociation. 
Instead, the phenomenology of soft hadronic processes based on Gribov--Regge theory  \cite{r1, r2} has been employed to describe these processes at high energies. 
Therefore, it is extremely important to constrain the phenomenological parameters based on measurement data to obtain a
correct understanding of the various diffractive processes and their accurate contribution to the total inelastic collisions. 

An adequate understanding of diffractive processes can help to improve Monte Carlo (MC) hadronic interaction models and event generators. 
Hadronic interaction models are widely used to simulate cosmic-ray interactions in the atmosphere. 
In experimental studies of high-energy cosmic rays (HECRs), the properties of primary HECRs are reconstructed from the measured characteristics of nuclear-electro-\ magnetic cascades induced in the atmosphere (so-called extensive air showers). 
Determining the primary mass composition and reconstructing the primary energy depend strongly on the MC procedures used for numerical simulations of air showers. 
Limitations in the modeling of hadronic interactions and the largely unknown model uncertainties lead to large uncertainties in interpreting the measurement data \cite{r3, r4}. 

% Fig. 1
\begin{figure*}
\centering
  \includegraphics[width=0.75\textwidth]{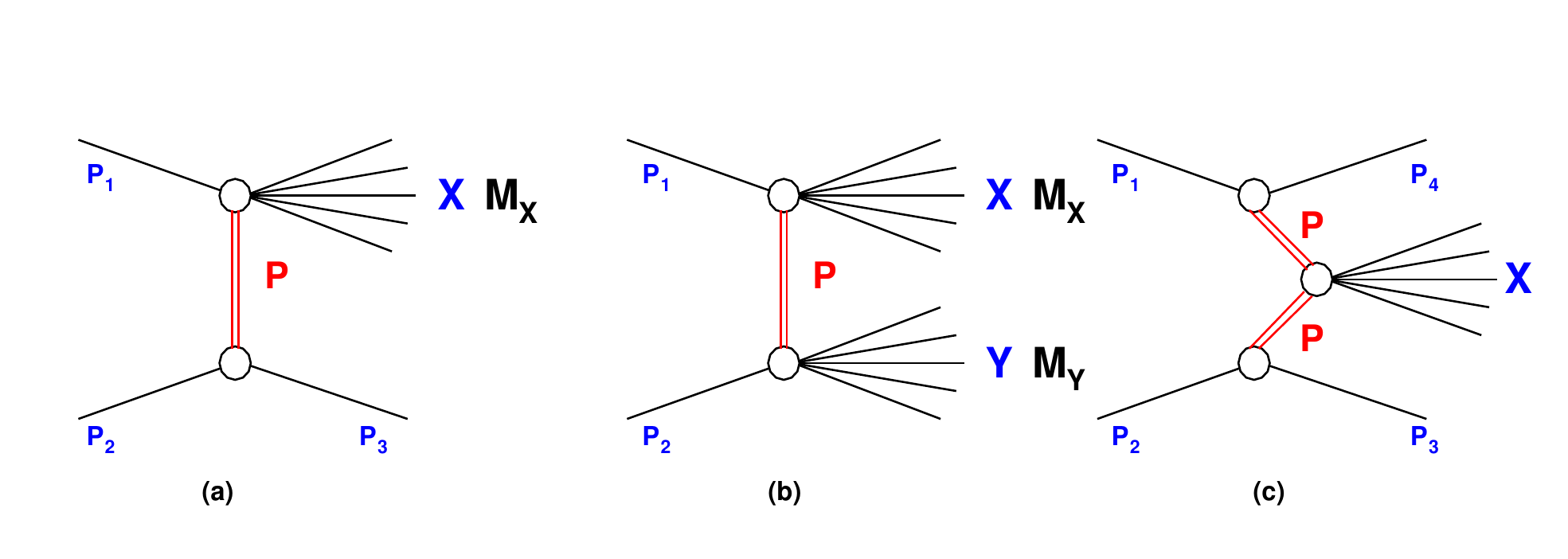}
\caption{Illustration of (a) single diffraction, (b) double diffraction, and (c) central diffraction with the pomeron exchanged in a proton-proton collision. $M_{X}$ and $M_{Y}$ are the invariant masses of the dissociated systems $X$ and $Y$.}
\label{Feynman}       
\end{figure*}

Partial cross sections of high-mass diffractive dissociations were measured by ATLAS \cite{r5, r6}, CMS \cite{r7, r8}, and ALICE \cite{r9} Collaborations at Large Hadron Collider (LHC) energies. 
In contrast, for obtaining the overall inelastic cross section, the cross section of low-mass diffractive dissociation is estimated by extrapolation based on MC simulations. 
The cross section of low-mass diffraction ($M_{X} < 3.4$ GeV) reported by the TOTEM collaboration, $2.62\pm 2.17$ mb \cite{r10}, at $\sqrt{s} = 7$ TeV explains the difficulty in this estimation. 
Though the total inelastic cross section is precisely measured by the TOTEM and ATLAS ALFA experiments using the Roman Pot technique \cite{r11, r12}, the cross section fractions among nondiffractive, high-mass diffractive, and low-mass diffractive dissociations is still an open question in the hadronic process.
Very forward (VF) detectors, covering zero-degree collision angles, have unique sensitivity to low-mass diffractive processes. 
Accordingly, applying the rapidity gap measurement based on central rapidity information makes it possible to access pure low-mass diffractive processes. 
Therefore, forward particle cross sections derived from such pure low-mass diffractive cases identified by central information can provide an opportunity for constraining hadronic interaction models. 

In this paper, parts of the ATLAS detector \cite{r12} and the LHCf detectors \cite{r13} located at interaction point 1 (IP1) of the LHC are considered to be representatives of the central detectors and VF detectors, respectively. 
The ATLAS inner detector (ID) measures particle momentum and vertex information with full azimuthal ($\phi$) and |$\eta$| < 2.5 pseudorapidity coverage.
For studies of minimum-bias measurements, this detector can provide information on charged tracks with a $p_{T}$ threshold as low as 100 MeV. 
The LHCf detectors were installed in the target neutral absorber (TAN) located $\pm$140 m from IP1. 
The detectors were designed to measure forward neutral particles (e.g., neutrons, photons, and $\pi^{0}$s) over a pseudorapidity range $|\eta| > 8.4$. 
The photon and hadron energy thresholds are 200 and 500 GeV, respectively. 
The ATLAS--LHCf common data acquisition experiment is dedicated to measuring and classifying diffractive dissociation. 
Since ATLAS and LHCf have totally different detector acceptances, this common experiment not only enhances the trigger efficiency for inelastic processes but also addresses some specific processes with each other's tagging information. 
    
In the present work, three subjects were investigated based on MC simulation. 
We first investigated the different contributions of nondiffractive and diffractive components to the forward neutral particle cross sections and the differences among models. 
Then, we evaluated the performance to identify the diffractive dissociation on the corresponding cross sections of neutral particles expected by the VF detector by applying a simple selection based on central detector information. 
Finally, we studied the sensitivity range in diffractive mass of the common experiment using VF and central detectors.
%=========================================================================================
%=========================================================================================
\section{Diffractive dissociation}
In high-energy proton interactions, Regge theory describes diffractive processes as a $t$-channel reaction, which is dominated by the exchange of an object with vacuum quantum numbers called {\em pomerons\/} \cite{r14, r15}. 
It is usually recognized that diffractive processes are composed of single-diffraction (SD; Fig.~\ref{Feynman}a ), double-diffraction (DD; Fig.~\ref{Feynman}b), and central-diffraction (CD; Fig.~\ref{Feynman}c) terms. 
An operational characteristic of diffractive interactions is the large angular separation between the final state systems called the rapidity gap $\Delta \eta$.
The size and location of $\Delta \eta$ in pseudorapidity phase space can be used to determine the type of diffraction. 
In the SD case, it is known that the relationship between the observables $\Delta \eta$ and $\xi_{X}$ is
\begin{equation}
\Delta  \eta \simeq -\ln(\xi_{X}),  
\end{equation} 
where $\xi_{X}=M_{X}^2/s$ with $\sqrt{s}$ being the total energy in the center-of-momentum frame.         
%=========================================================================================
%=========================================================================================
%% Fig. 2      
\begin{figure}
  \includegraphics[width=0.50\textwidth]{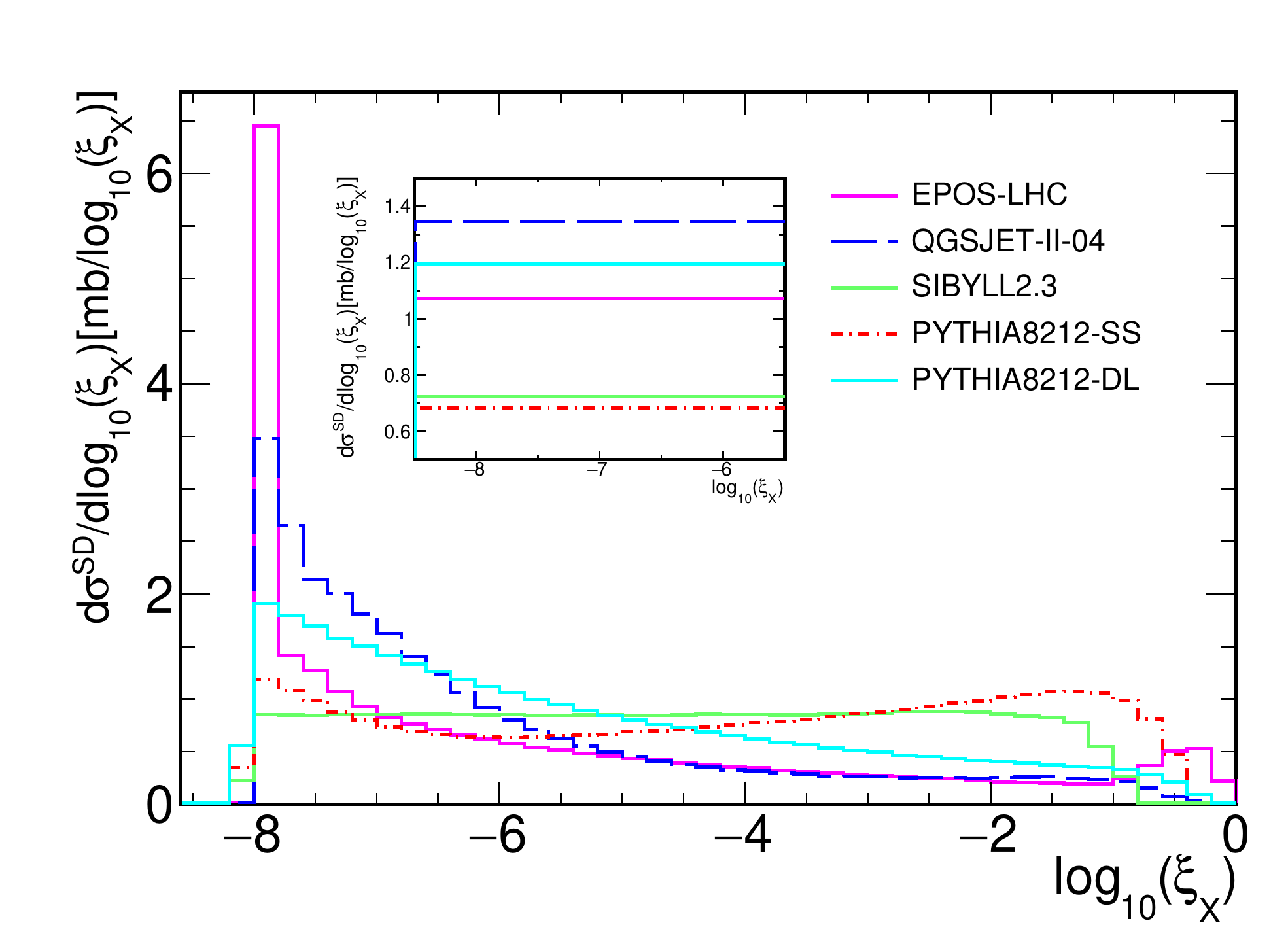}
\caption{SD ($pp \rightarrow pX$; blue) cross section shown as a function of $\log_{10}\xi_{X}$. MC predictions with EPOS-LHC (magenta), QGSJET-II-04 (blue dashed), SIBYLL2.3 (green), PYTHIA8212-SS (red dotted-dashed), and PYTHIA8212-DL (cyan) compared with each other. The comparison of low-$M_{X}$ SD cross section predicted by models is shown in the inset.}
\label{CroSecModel} 
\end{figure} 

%% Fig. 3 
\begin{figure*}
  \includegraphics[width=1.04\textwidth]{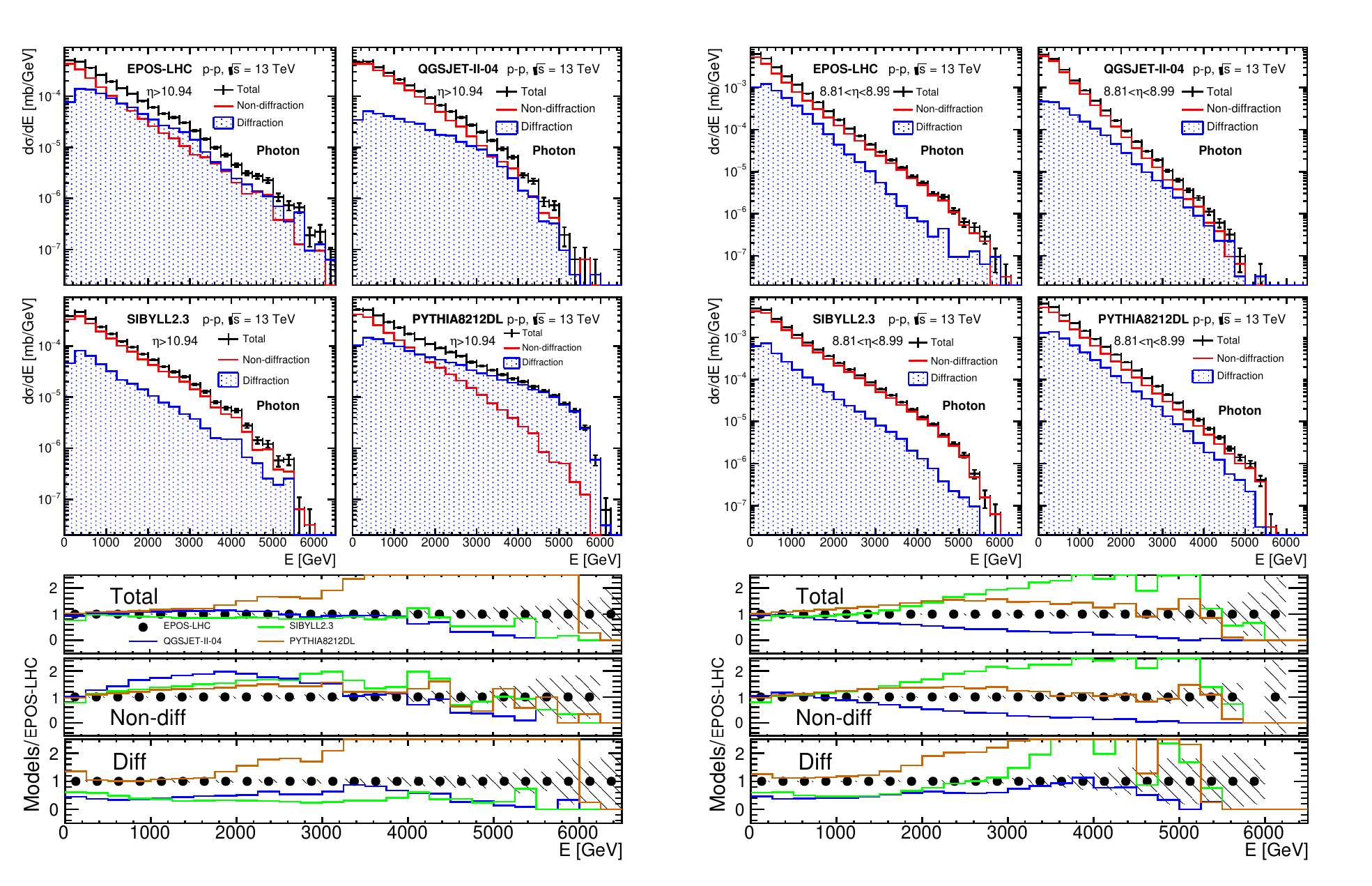}
\caption{Photon spectra at $\eta > 10.94$ (left) and $8.81< \eta <8.99$ (right) (top four panels in each set). These are generated by EPOS-LHC, QGSJET-I\hspace{-.1em}I-04, SYBILL 2.3, and PYTHIA 8212DL, respectively. The total photon spectra (black) were classified by nondiffraction (red) and diffraction (blue) according to MC true flags. The bottom three plots show the ratios of the spectra of EPOS-LHC (black markers), QGSJET-I\hspace{-.1em}I-04 (blue lines), SYBILL 2.3 (green lines), and PYTHIA8212DL (orange lines) to the spectrum of EPOS-LHC. The top, middle, and bottom plots correspond to total, nondiffraction, and diffraction, respectively.}
\label{TrueSpecG}       
\end{figure*}
%% Fig. 4
\begin{figure*}
  \includegraphics[width=1.04\textwidth]{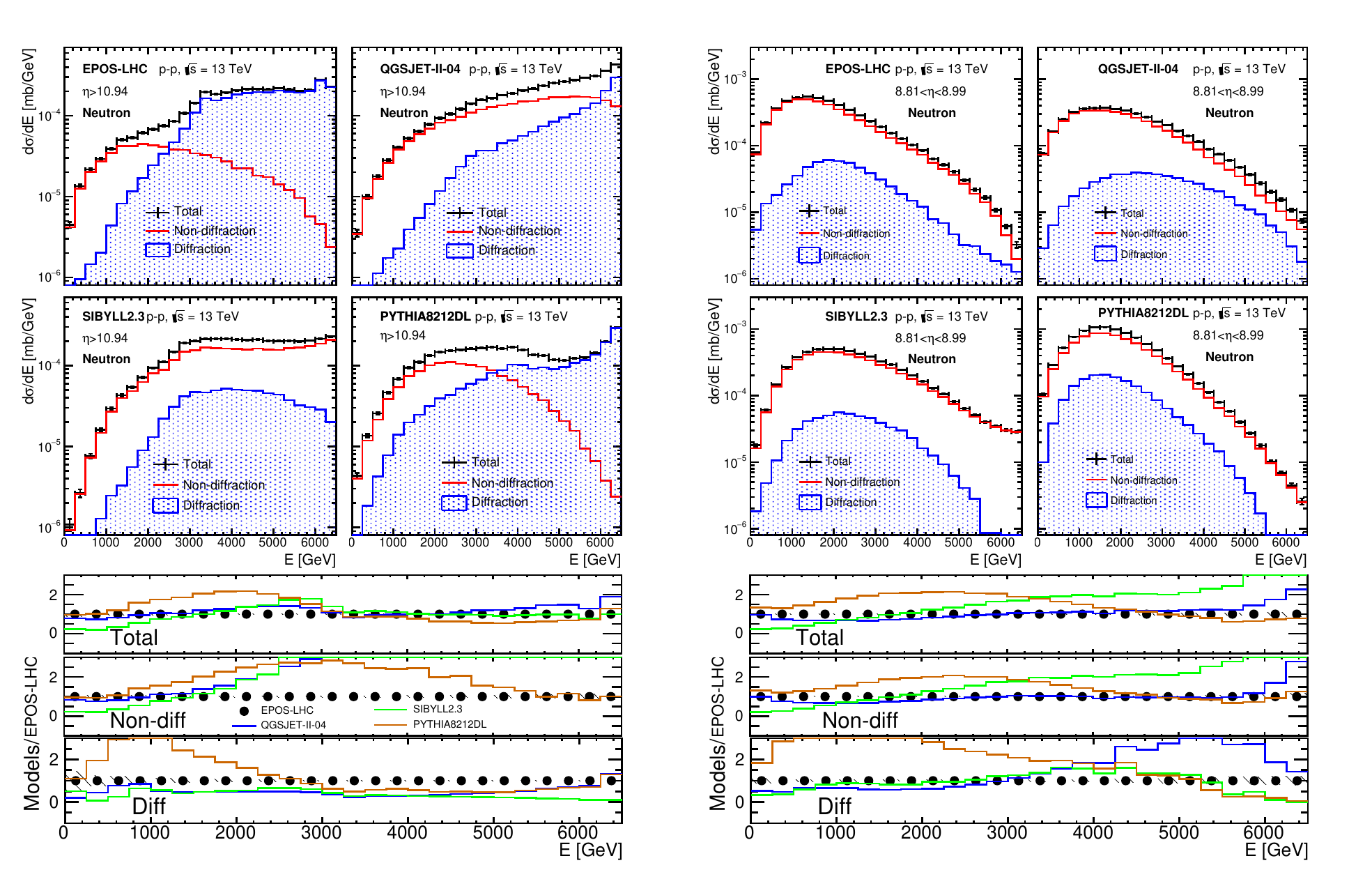}
\caption{Neutron spectra at $\eta > 10.94$ (left) and $8.81<\eta<8.99$ (right) (top four panels in each set). These are generated by EPOS-LHC, QGSJET-I\hspace{-.1em}I-04, SYBILL 2.3, and PYTHIA 8212DL, respectively. The total neutron spectra (black) were classified by nondiffraction (red) and diffraction (blue) according to MC true flags. The bottom three plots show the ratios of the spectrum of EPOS-LHC (black markers), QGSJET-I\hspace{-.1em}I-04 (blue lines), SYBILL 2.3 (green lines), and PYTHIA8212DL (orange lines) to the spectrum of EPOS-LHC. The top, middle, and bottom plots correspond to total, nondiffraction, and diffraction, respectively.}
\label{TrueSpecN}       
\end{figure*}

%% Fig. 5
\begin{figure*}
  \includegraphics[width=1.04\textwidth]{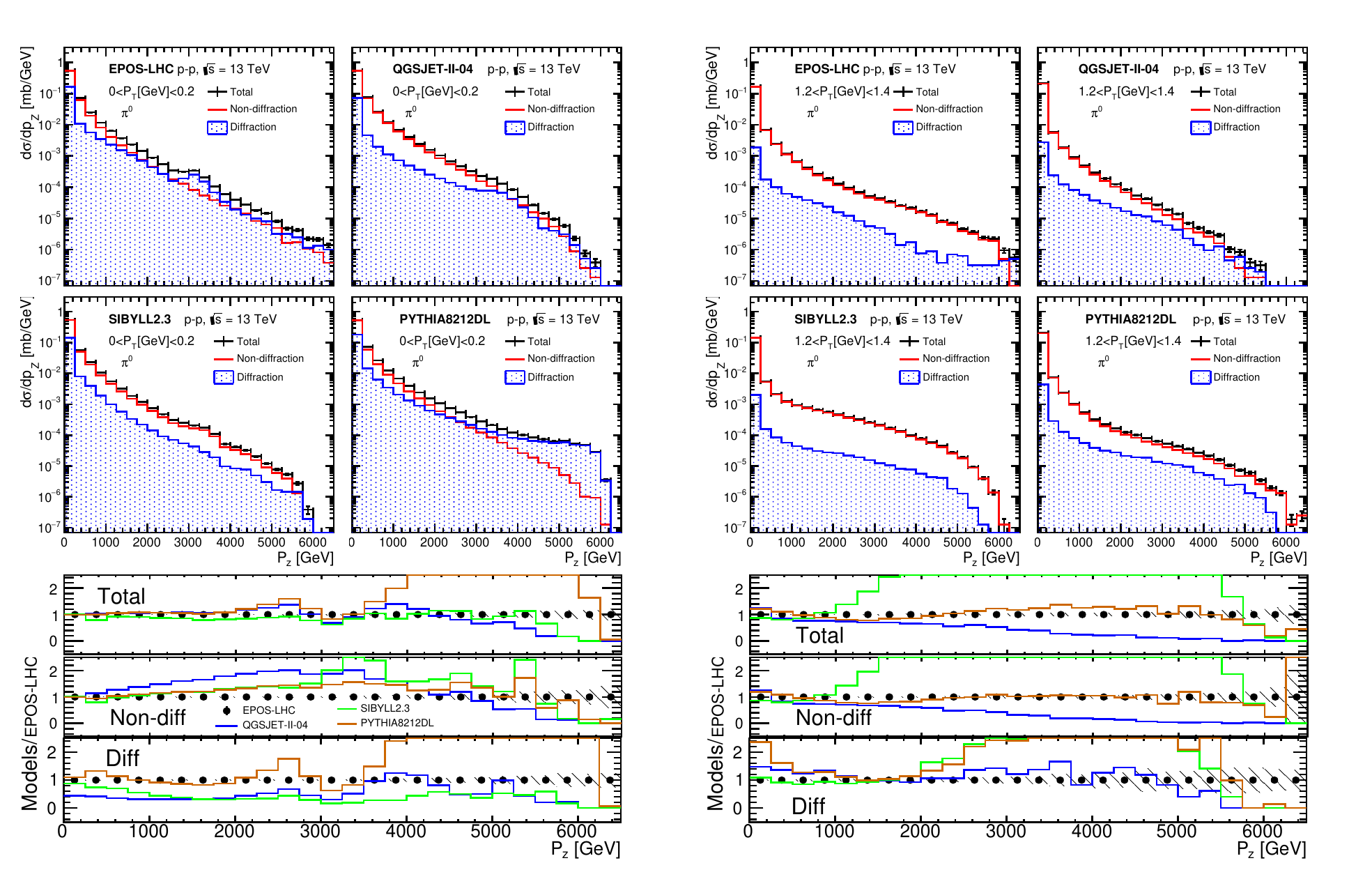}
\caption{$\pi^{0}$ spectra shown with $0 < P_{T} < 0.2$ GeV (left) and $1.2 < P_{T} < 1.4$ GeV (right) for comparison. In each $P_{T}$ phase region, the top four panels show $\pi^{0}$ spectra generated by EPOS-LHC, QGSJET-I\hspace{-.1em}I-04, SYBILL 2.3, and PYTHIA 8212DL, respectively. The total $\pi^{0}$ spectra (black) were classified by nondiffraction (red) and diffraction (blue) according to MC true flags. The bottom three plots show the ratios of the spectrum of EPOS-LHC (black markers), QGSJET-I\hspace{-.1em}I-04 (blue lines), SYBILL 2.3 (green lines), and PYTHIA8212DL (orange lines) to the spectrum of EPOS-LHC. The top, middle, and bottom plots correspond to total, nondiffraction, and diffraction, respectively.}
\label{TrueSpecPi}       
\end{figure*}
%%%
%=========================================================================================
%=========================================================================================
\section{Monte Carlo simulation}
In this analysis, MC simulation samples were produced using four interaction models for comparison. 
$p$--$p$ collision events at $\sqrt{s} = 13$ TeV were simulated by each model, and trigger conditions of a VF detector were applied based on the energy, particle type, and $\eta$ according to the LHCf case.
Four MC generators are extensively used in cosmic-ray observations and high-energy experiments: EPOS-LHC \cite{r16}, QGSJET-I\hspace{-.1em}I-04 \cite{r17}, SYBILL 2.3 \cite{r18, r19}, and PYTHIA 8212 \cite{r20, r21}.
All these models are post-LHC generators tuned by using the LHC Run1 data. 
The first three simulation samples were generated by using the integrated interface tool CRMC v1.6.0 \cite{r22}, whereas for PHYHIA, its own front-end was used.  

For the PYTHIA8 generator, Monash event tuning \cite{r23} was employed in this analysis. 
Minimum-bias data and underlying event data from the LHC were used for constraining the parameters. 
The new NNPDF2.3 LO PDF set was adopted in the event tuning. 
By default, PYTHIA8 uses the Schuler and Sj$\rm \ddot{o}$strand (SS) parameterization \cite{r24} of the pomeron flux. 
In addition, an alternative pomeron flux model, the Donnachie and Landshoff (DL) \cite{r25} model, with a linear pomeron trajectory $\alpha_{\mathbb{P}}(t) = 1+\Delta +\alpha^{'} t$ is also implemented. 
The default value of variable parameters $\Delta$ and $\alpha^{'}$ are 0.085 and 0.25 GeV$^{-2}$ \cite{r26}, respectively.
According to the ATLAS minimum-bias measurement in $p$--$p$ collisions at $\sqrt{s}=13$ TeV, the PYTHIA8212DL model gives the best description of the number of hits detected by the minimum-bias trigger scintillators \cite{r27}. 
Therefore, the PYTHIA8212DL model was employed in this analysis. 
The total inelastic cross sections in $p$--$p$ collisions at $\sqrt{s} = 13$ TeV implemented in each model were 78.984, 80.167, 78.420, and 79.865 mb, corresponding to EPOS-LHC, QGSJET-II-04, PYTHIA8212DL, and SIBYLL2.3, respectively.

Given the model differences in the treatments of diffractive components, not only the predicted diffraction cross sections but also the diffractive mass distributions are important. 
Figure ~\ref{CroSecModel} shows the SD ($pp \rightarrow pX$) cross sections in each $\xi_{X}$ interval predicted by several models. The different spectral shapes come from the different approaches to the diffraction treatment implemented in the models.
There are large differences among models in both the high and low diffractive mass regions. 
The flat SD cross section distribution of the SIBYLL2.3 model corresponds to a diffractive mass distribution described as $dM_{X}^{2}/M_{X}^2$ \cite{r26}. 
The PYTHIA8212 model (SS pomeron flux) uses a treatment similar to that of SIBYLL2.3 for the diffractive mass distribution \cite{r24}. 
In the high diffractive mass regions, EPOS-LHC tuned $d\sigma^{SD}/d\Delta \eta$ \cite{r16} by comparing with the data of the ATLAS rapidity gap distribution shown in \cite{r6}.
The inset of Fig.~\ref{CroSecModel} shows the low-mass SD cross sections of each model in the $\xi_{X}$ interval $-8.5 < \log_{10}(\xi_{x}) < -5.5$. 
The QGSJET-II-04 model uses different transverse profiles for pomeron emission vertices by different elastic scattering eigenstates \cite{r17, r28}. 
This leads to the larger cross sections in the low-mass regions at very high energies.
%=========================================================================================
%=========================================================================================
\section{Diffractive and nondiffractive contributions to the VF photon, neutron, and $\pi^0$ spectra}
The LHCf collaboration has published several forward neutral particle spectra at different collision energies, but no hadronic interaction model can describe the LHCf results perfectly \cite{r29, r30, r31, r32}. 
To address the origin of the differences between the experimental data and the model predictions, separating the VF-triggered events to diffractive and nondiffractive contributions is important. 

In this analysis, all the events from each simulation are classified into nondiffractive and diffractive collisions by using MC flags, where the SD, DD, and CD events are together treated as diffraction. 
It is noted that the SIBYLL2.3 model does not implement CD processes. 
The simulated VF photon and neutron spectra are shown in the top four panels of Figs. ~\ref{TrueSpecG} and ~\ref{TrueSpecN} for two fiducial areas, |$\eta|>10.94$ (left) and $8.81<\eta<8.99$ (right), respectively. 
Meanwhile, Fig.~\ref{TrueSpecPi} shows the $\pi^{0}$ $p_{z}$ spectra at the fiducial $p_{T}$ phase spaces of $0<p_{T}<0.2$ GeV (left) and $1.2<P_{T}<1.4$ GeV (right), respectively. 
The spectra of total, nondiffractive, and diffractive components of four MC samples are compared with each other. 
In the bottom three panels of Figs. ~\ref{TrueSpecG}, \ref{TrueSpecN}, and \ref{TrueSpecPi}, the ratios of the spectra divided by the EPOS-LHC results are plotted separately for total, nondiffractive, and diffractive components. 
Clearly, the nondiffraction and diffraction implemented in each model are very different. 
Especially, PYTHIA8212DL predicts the largest diffractive contribution at high photon energies at $|\eta|>10.94$ and in the $\pi^{0}$ $p_{z}$ spectrum at $0<p_{T}<0.2$ GeV. 
There is no large difference between models of the neutron total spectra at $|\eta|>10.94$. 
In contrast, comparing the individual contribution of nondiffractive and diffractive components, one sees that the neutron spectra of EPOS-LHC and PYTHIA8212DL are dominated by diffraction, but those of QGSJET-II-04 and SIBYLL2.3 are dominated by nondiffraction at high energies. 
As shown in Fig.~\ref{TrueSpecPi}, SIBYLL2.3 predicts a larger contribution for all components at $1.2<p_{T}<1.4$ GeV. 
It is also found that the larger the value of $p_{T}$ is, the larger is the contribution from all components predicted by SIBYLL2.3.
Additionally, in Figs.~\ref{TrueSpecN} and ~\ref{TrueSpecPi}, neutron and $\pi^{0}$ spectra of EPOS-LHC and QGSJET-II-04 exhibit a bump or kink structure at $\sim$3 TeV. 
In the EPOS-LHC model, this structure is due to the simple approach used for the pion-exchange process, whereas in QGSJET-II-04 it is due to the selected kinematics, the contribution to which is mainly from low-mass diffraction. 
%=========================================================================================
%=========================================================================================
\section{Identification of diffraction with central track information}
Because of the large differences found among different hadronic interaction models, it is important to classify the observed VF spectra into nondiffraction or diffraction by using experimental data. 
Although, in principle, diffractive collisions can be identified by measuring the rapidity gap of the final state,
it is experimentally difficult to measure rapidity gaps preciously because of the limited pseudorapidity coverage and energy threshold of the detectors. 
However, improved experimental techniques have helped in reaching lower $p_{T}$ thresholds and larger rapidity ranges. 
The results from measurements of rapidity gaps over limited pseudorapidity ranges have been reported by ATLAS \cite{r6}, CMS \cite{r8}, and ALICE \cite{r9} Collaborations. 
Similarly, such rapidity gap techniques can be adopted for diffractive event identification.
%=========================================================================================
\subsection{Diffraction selection criteria}
%% Table 2
\begin{table}
\centering
\caption{Efficiency and purity of central-veto selection with different track conditions.}
\label{effpur}
\begin{tabular*}{\columnwidth}{@{\extracolsep{\fill}}lllll@{}}
\hline \hline
Parameter &  $N_{track}=0$  & $N_{track}\leq 1$  & $N_{track}\leq 2$  &$N_{track}\leq 5$\\
\hline
Efficiency ($\epsilon$)            & 0.493     & 0.556     & 0.619    &0.691    \\
Purity (p)                                 & 0.995     & 0.991     & 0.982    &0.950    \\
\hline \hline
\end{tabular*}
\end{table}
%% Fig. 6
\begin{figure}
  \includegraphics[width=0.50\textwidth]{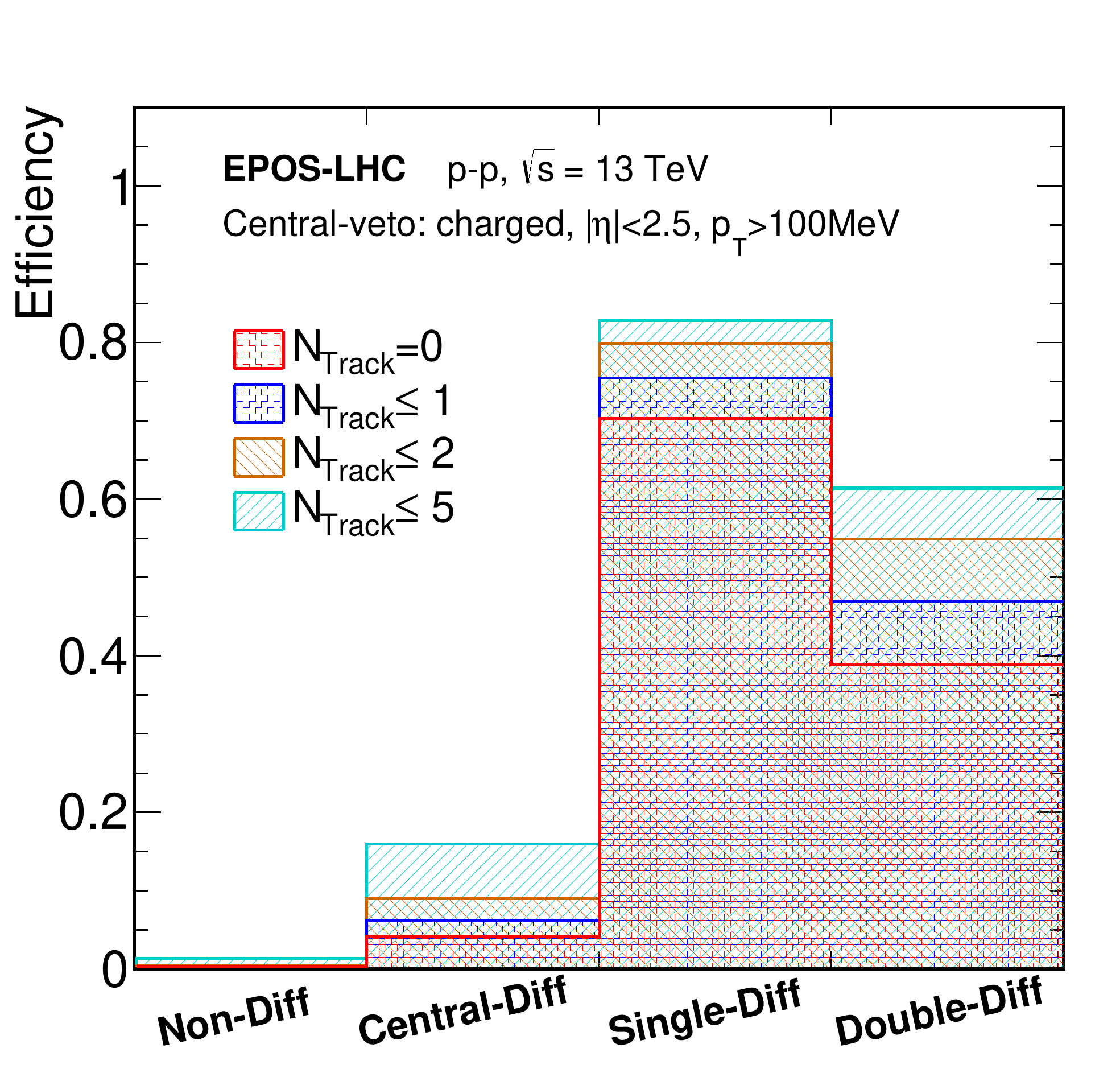}
\caption{Diffraction selection efficiency with different central-veto selection conditions: $N_{track} = 0$ (red), $N_{track} \leq   1$ (blue), $N_{track} \leq   2$ (brown), and $N_{track} \leq   5$ (cyan) charged particles at $|\eta| < 2.5$ with $p_{T} > 100$ MeV.}
\label{TypeEff}      
\end{figure}
The identification of the type of diffraction requires detection of a large rapidity gap because small rapidity gaps may be produced by fluctuations in nondiffractive particle production \cite{r33}.
Consequently, a small number of particles is expected in the central detector, for instance, the ATLAS detector. 
If an event has a small number of tracks, $N_{track}$, it is more likely to be a diffractive event. 
This is the basic idea in this analysis used to identify diffractive events.
In other words, having a small number of charged tracks in the central region is used to {\it veto} nondiffractive events.
It is assumed that the central detector can count $N_{track}$ with $p_{T}>100$ MeV at $|\eta|<2.5$. 
The performance of central-veto event selection was studied for different criteria of $N_{track}$, $N_{track}=0$, $N_{track}\leq 1$, $N_{track}\leq 2$, and $N_{track}\leq 5$ in \cite{r34}.
If the event survives central-veto selection, it is classified as a diffractive-like event; otherwise, it is classified as a nondiffractive-like event. 
According to MC true flags, events can be classified as nondiffraction (ND), CD, SD, and DD. 
By applying central-veto selection to each event, the selection efficiency ($\epsilon$) and purity ($\kappa$) of diffractive event selection are defined as 
\begin{eqnarray}
\epsilon = \frac{(N_{CD}+N_{SD}+N_{DD})_{central \ veto}}{N_{CD}+N_{SD}+N_{DD}}, \label{eff} \\
\kappa = \frac{(N_{CD}+N_{SD}+N_{DD})_{central \ veto}}{(N_{ND}+N_{CD}+N_{SD}+N_{DD})_{central \ veto}},\label{pur}
\end{eqnarray}
where $N_{ND}$, $N_{CD}$, $M_{SD}$, and $N_{DD}$ indicate the number of events triggered by a VF detector in each event category.
The suffix ${central \ veto}$ signifies number of events after applying central-veto event selection.

Figure~\ref{TypeEff} shows a comparison of the central-veto selection efficiency with the four criteria, which are calculated by using the EPOS-LHC simulation samples. 
It is clear that the efficiency rises as the $N_{track}$ threshold increases.
SD selection efficiency, for instance, increases from about 70\% to 80\% as $N_{track}=0$ changes to $N_{track}\leq 5$. 
The efficiency and purity of the central-veto selection for the four criteria are summarized in Table~\ref{effpur}. 
High selection purity (99.5\%) is achieved when the criterion is $N_{track}=0$ while it decreases only by 5\% when $N_{track}\leq 5$ is applied.
To aid our discussion using a simple analysis, we adopt the following criterion for the central veto (diffraction selection):
{\it There are no charged particles ($N_{track}=0$) in the kinematic range $|\eta| <2.5$ and $p_T>100$ MeV}.
%=========================================================================================
%%% Fig. 7
\begin{figure}
  \includegraphics[width=0.50\textwidth]{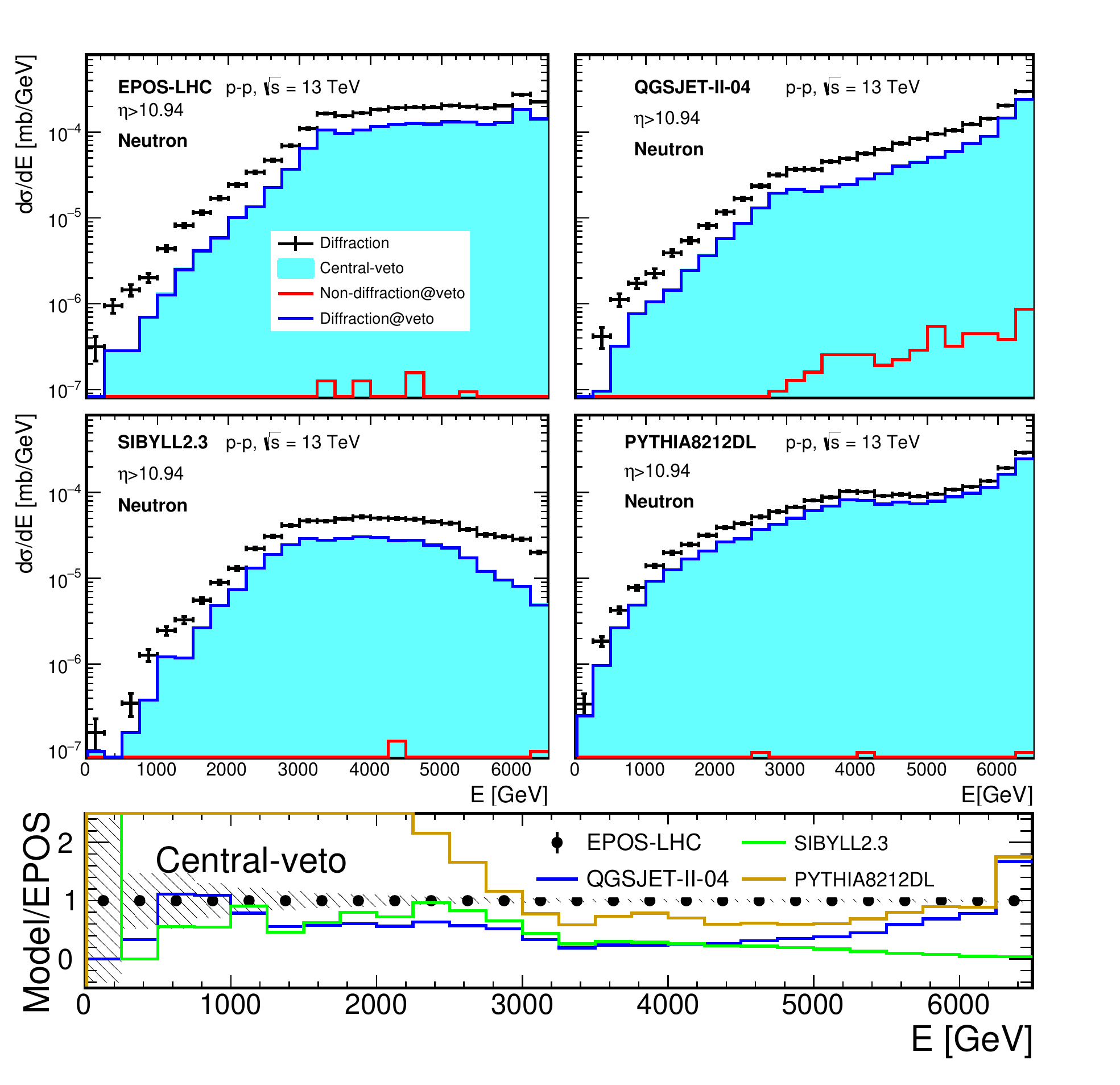}
\caption{Neutron spectra at $\eta > 10.94$ generated by EPOS-LHC, QGSJET-I\hspace{-.1em} I-04, SYBILL 2.3, and PYTHIA 8212DL. The top four panels show the spectra of true diffraction (black lines) and diffractive-like events corresponding to central-veto selection (filled gray areas), which are defined as events without any $P_{T} > 100$ MeV charged particles at $|\eta| < 2.5$; in addition, the central-veto events were classified by nondiffraction (red) and diffraction (blue) again according to MC true information. The bottom plot shows the ratios of the central-veto spectrum of each model to the central-veto spectrum of EPOS-LHC.}
\label{SelSpecN0}       
\end{figure}
%%% Fig. 8
\begin{figure}
  \includegraphics[width=0.50\textwidth]{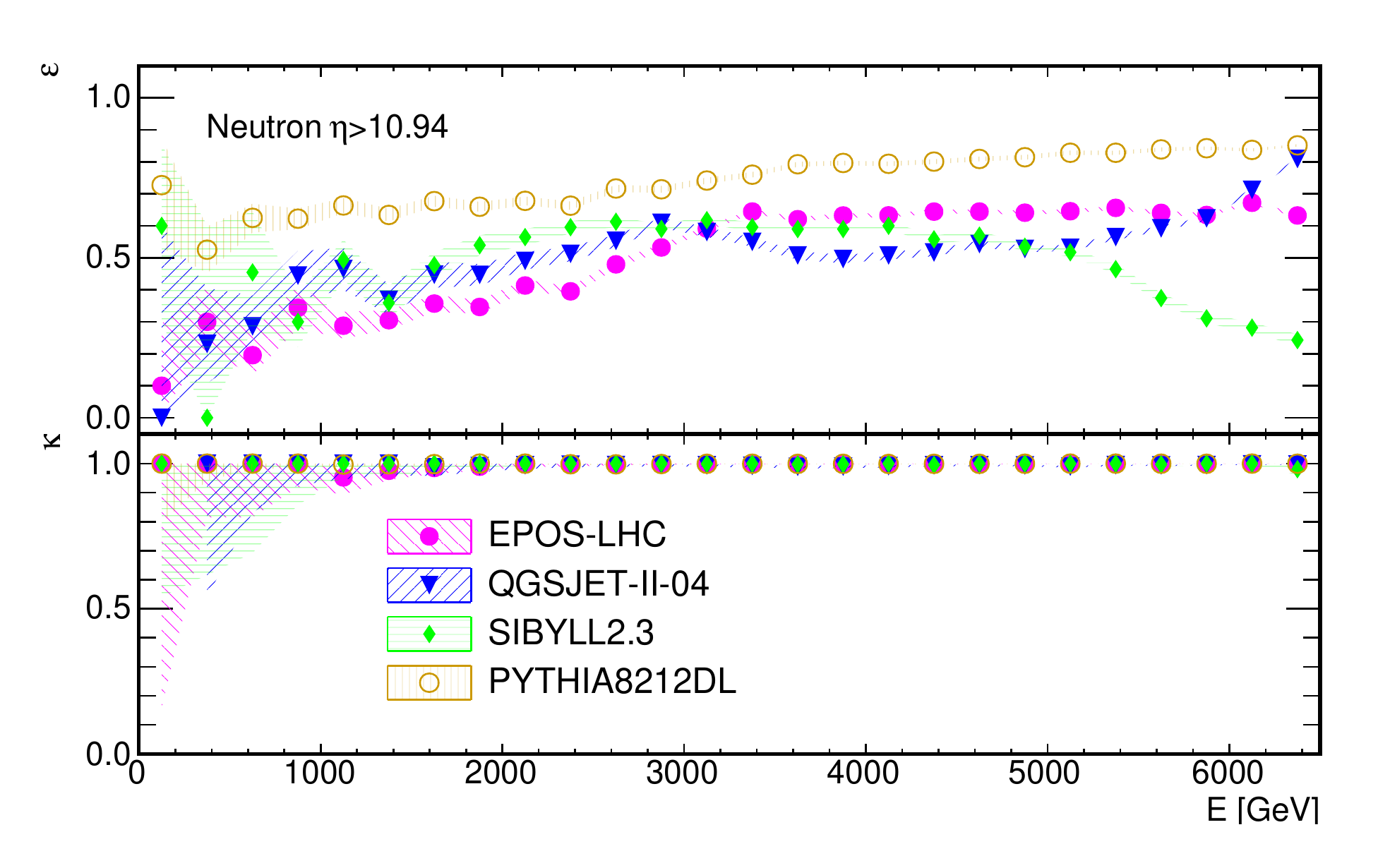}
\caption{Efficiency (top) and purity (bottom) of diffraction selection for the VF ($\eta > 10.94$) neutron spectra obtained by using the central-veto technique.}
\label{EffPurN0}       
\end{figure}

\subsection{Performance of central-veto selection}
\begin{figure}
  \includegraphics[width=0.50\textwidth]{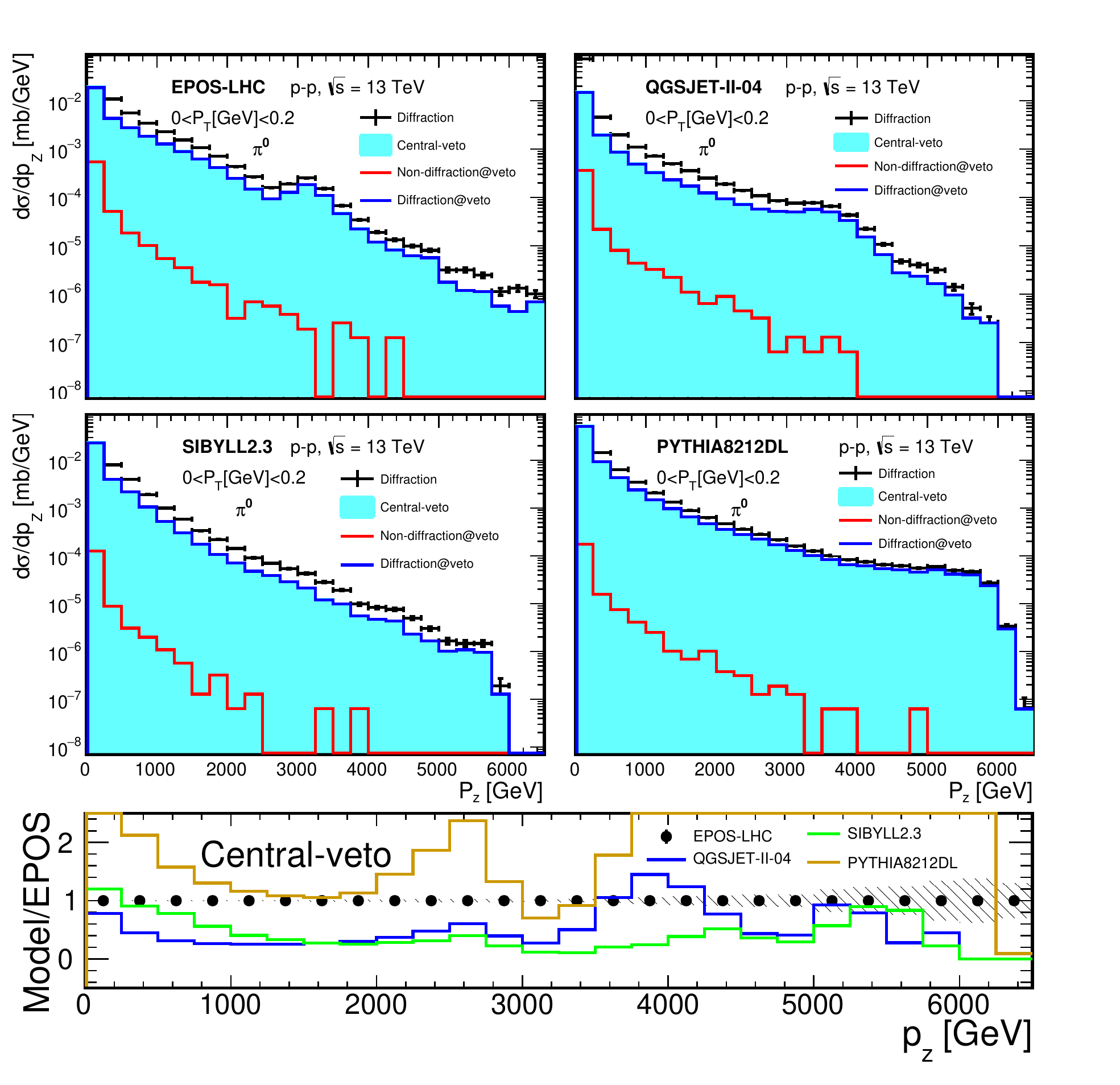}
\caption{$\pi^{0}$ spectra at $0 < p_{T}< 0.2$ GeV generated by EPOS-LHC, QGSJET-I\hspace{-.1em} I-04, SYBILL 2.3, and PYTHIA 8212DL. The top four panels show the spectra of true diffraction (black lines) and diffractive-like events corresponding to central-veto selection (filled gray areas), which are defined as events without any $p_{T} > 100$ MeV charged particles at $|\eta| < 2.5$; in addition, the central-veto events were classified by nondiffraction (red) and diffraction (blue) again according to MC true information. The bottom plots show the ratios of the central-veto spectrum of each model to the central-veto spectrum of EPOS-LHC.}
\label{SelSpecPi0}      
\end{figure}

\begin{figure}
  \includegraphics[width=0.50\textwidth]{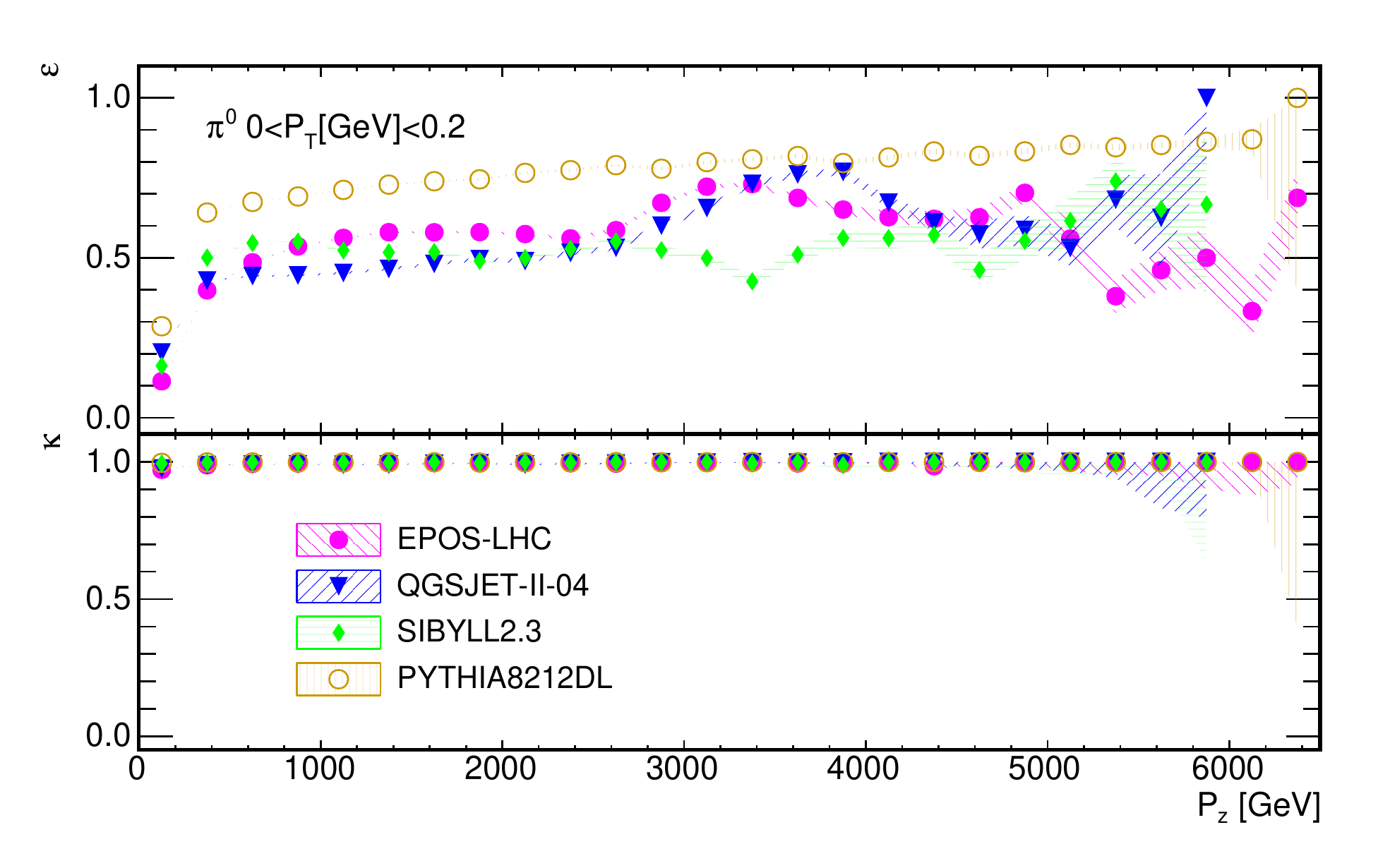}
\caption{Efficiency (top) and purity (bottom) of diffraction selection for the VF ($0 < p_{T}< 0.2$ GeV) $\pi^{0}$ $p_{z}$ spectra obtained by using the central-veto technique.}
\label{EffPurPi0}       
\end{figure}

To evaluate the performance of central-veto selection based on the VF spectra, the VF spectra were classified as nondiffractive-like and diffractive-like. 
A comparison of the VF neutron and $\pi^{0}$ spectra in the VF regions is shown in Figs. ~\ref{SelSpecN0} and ~\ref{SelSpecPi0}, respectively. 
They indicate that the spectra corresponding to events surviving central-veto selection keep almost the same shapes as the VF true diffractive spectra. 
Moreover, the number of misidentified diffractive-like events is very small, as shown by the red histograms.
Comparisons of the differential cross sections of surviving events from central-veto selection are shown in the bottom plots of Figs. ~\ref{EffPurN0} and ~\ref{EffPurPi0}. 
The differences among models are expected to be constrained directly by using experimental data.
The efficiency and purity of central-veto selection as function of energy were calculated with Eq. \ref{eff} and Eq. \ref{pur}, as shown in Figs. ~\ref{EffPurN0} and ~\ref{EffPurPi0}. 
It is clear that selection purity stays constantly high (at $\approx$ 100\%), independent of particle type, energy, and MC simulation model, 
whereas selection efficiency has a tendency to increase with increasing energy. 
In contrast from selection purity, selection efficiency exhibits differences among MC simulation models.
In particular, the bump structure in EPOS-LHC and QGSJET-II-04 mentioned above still remains on the efficiency spectra. 
In such a case, comparing measured data with the MC samples as shown in Figs. ~\ref{SelSpecN0} and ~\ref{SelSpecPi0} can not only constrain the diffraction cross sections in the VF region but also help in identifying the inherent problems in the model. 
%%% Fig. 9
%%% Fig. 10
%=========================================================================================
\subsection{Low-mass diffraction}
The high-mass diffraction cross sections $d\sigma^{SD}/d\Delta  \eta$ at LHC energies were measured by ATLAS \cite{r5, r6}, CMS \cite{r7, r8}, and ALICE \cite{r9}. 
Typically, owing to the limited acceptance of these detectors, the rapidity gap signatures of events at around $-6 < \log_{10}(\xi_{x}) < -2$ can be identified in the case of ATLAS; these correspond to the lower and upper limits of $M_{X}$ of $\sim$13 and 1300 GeV at $\sqrt s = 13$ TeV, respectively. 
This fiducial region excludes the measurement of low-mass diffraction for determination of the total inelastic cross sections. 
As mentioned, low-mass diffraction is the main source of systematic uncertainties \cite{r5, r7} in the determination of inelastic cross sections. 

\begin{figure}
  \includegraphics[width=0.50\textwidth]{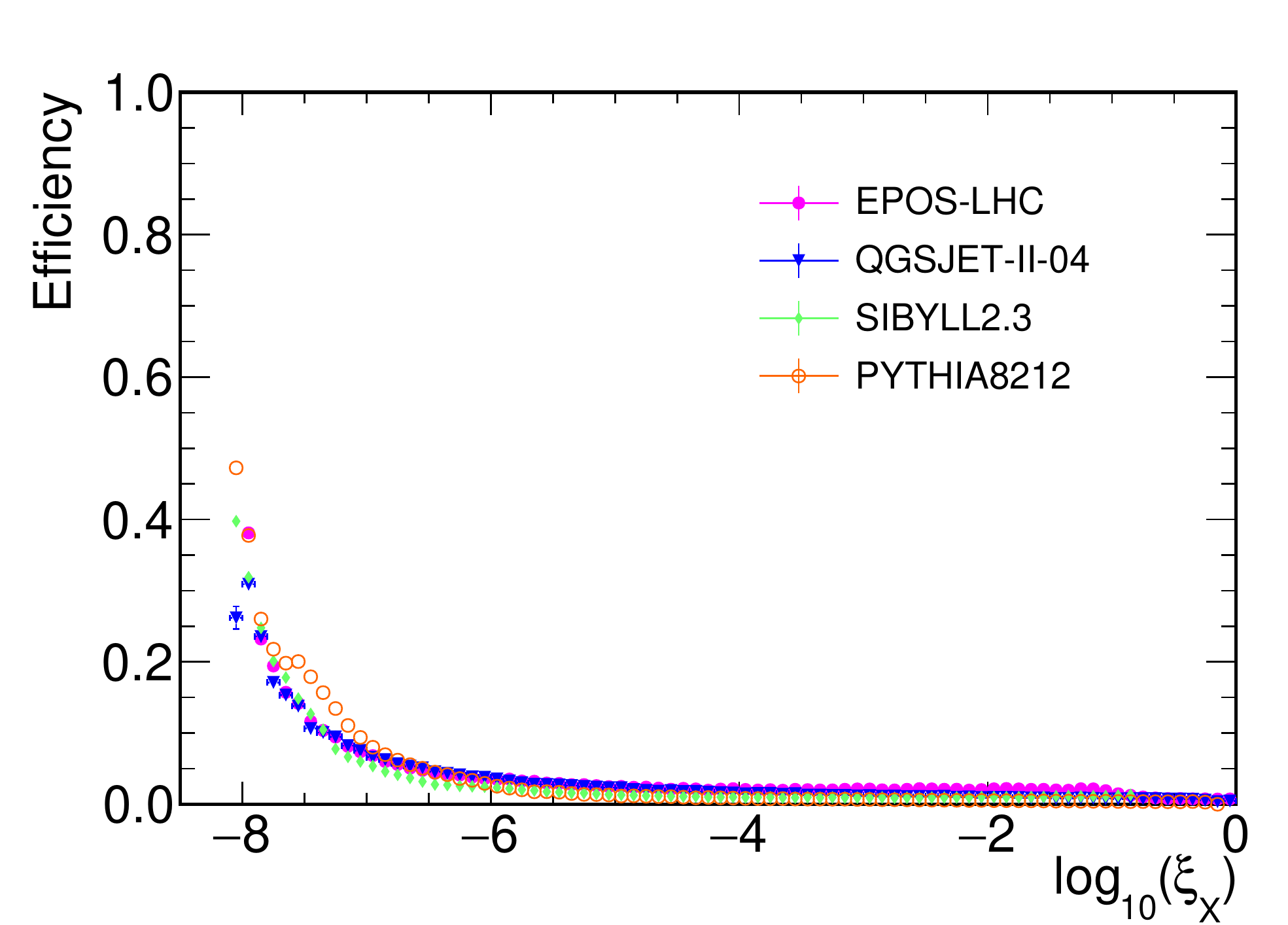}
\caption{LHCf detection efficiency as a function of $\log_{10}\xi_{X}$, which is simulated by four MC simulation samples. The trigger conditions for LHCf detectors at $\sqrt{s}=13$ TeV are $E_{\gamma}>200$ GeV and $E_{h}>500$ GeV. Only the SD ($pp \rightarrow pX$) component is used for this calculation.} 
\label{TrigEff} 
\end{figure}

Roman Pot detectors and VF detectors have sensitivities to low-mass diffractive processes. 
To evaluate the performance of VF detectors for the detection of low-mass diffraction, the LHCf detector is considered as representative. 
The acceptances of the LHCf detector for the forward neutral particles predicted by MC interaction models are shown in Fig.~\ref{TrigEff}. 
In the region of $\log_{10}(\xi_{x}) > -5.5$, the SD detection efficiency of the LHCf detector is only a few percent. 
The detection efficiency, however, increases below $\approx$$\log_{10}(\xi_{x}) = -6$ and reaches a maximum of $\sim$40\% at $\log_{10}(\xi_{x}) = -8$. 
In contrast, central detectors exhibit a totally opposite tendency of detection efficiency. 
For instance, the ATLAS detector has almost 100\% SD detection efficiency in the region of $\log_{10}(\xi_{x}) > -5$ but decreases rapidly to 0 at $\log_{10}(\xi_{x}) = -7$ \cite{r27}. 
Therefore, the common experiment using central and VF detectors can enhance detection efficiency, especially for low-mass processes.       

\begin{figure}
  \includegraphics[width=0.50\textwidth]{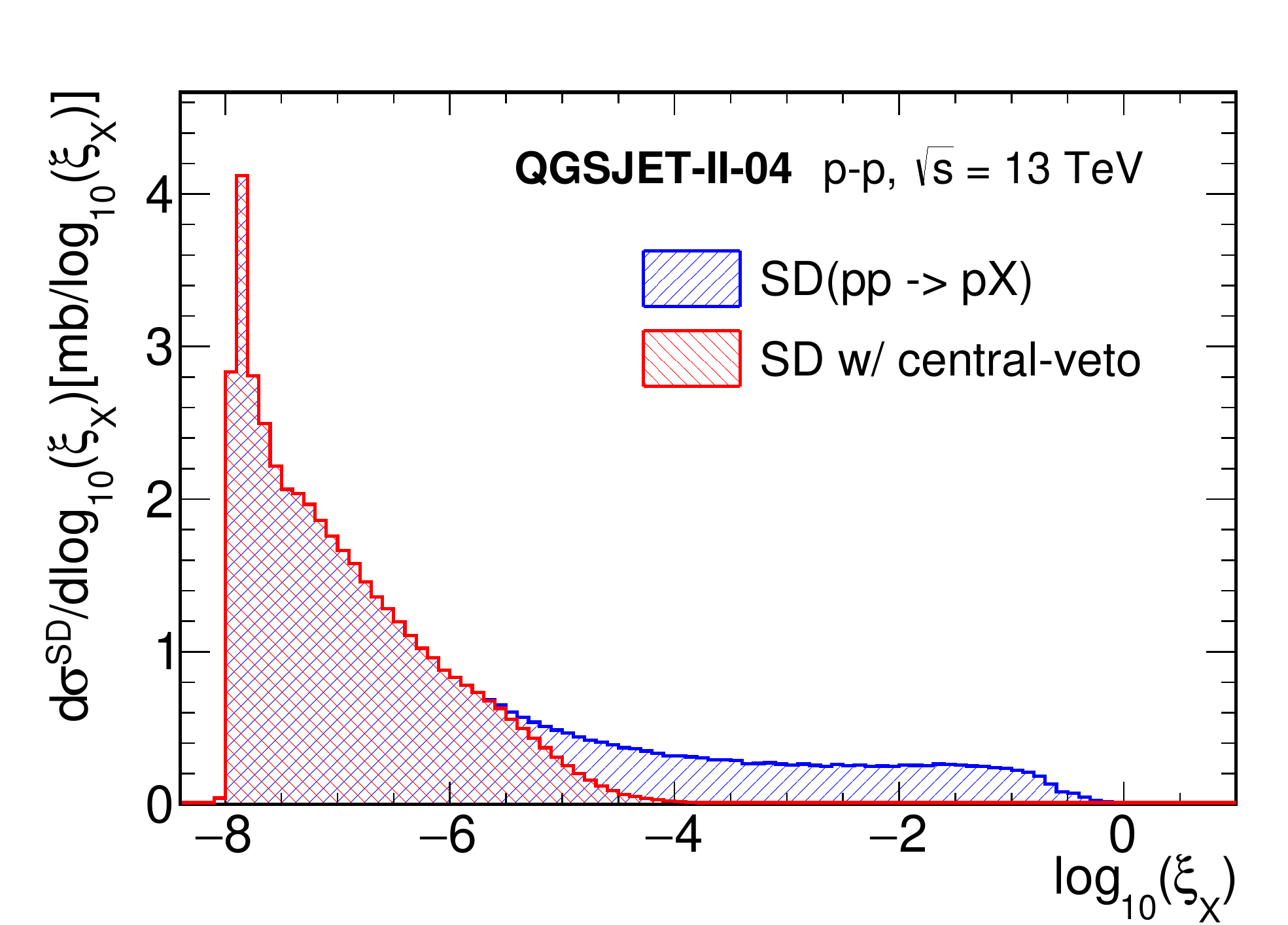}
\caption{SD ($pp \rightarrow pX$; blue) cross section as a function of $\log_{10}\xi_{X}$ predicted by using QGSJET-II-04 MC samples. This is compared with the SD cross section after applying the central-veto selection (red).}
\label{CroSec} 
\end{figure}

According to QGSJET-II-04 simulation predictions, most of the events survived from the central-veto selection are from the low-mass diffraction as shown in Fig.~\ref{CroSec}. 
In particular, all the low-mass diffractive events at $\log_{10}(\xi_{x}) < -5.5$ detected by VF detector survived from the central-veto selection, whereas all the high-mass diffractive events at $\log_{10}(\xi_{x}) > -4$ were excluded. 
In the other word, the filled histogram in Figs. ~\ref{SelSpecN0} and ~\ref{SelSpecPi0} are mostly derived from the low-mass diffractive processes at $\log_{10}(\xi_{x}) < -5.5$. 
Therefore, the common experiment using VF and central detectors can provide a chance to verify the results of low-mass diffraction reported by TOTEM \cite{r10} and impose a constraint on the treatment of low-mass diffraction implemented in MC simulation models through VF neutral particle spectra.       
%% Fig. 11
%% Fig. 12
%
%=========================================================================================
%=========================================================================================
\section{Conclusion}
We studied the nondiffractive and diffractive contributions to VF particle production using MC predictions in $p$--$p$ collisions at $\sqrt s = 13$ TeV. 
For the forward photon and $\pi^0$ energy spectra, PYTHIA8212DL predicts the largest diffractive contributions at high energies. 
In the cases of neutron differential cross sections at high energies, EPOS-LHC and PYTHIA8212DL are dominated by diffraction at $|\eta|>10.94$ while QGSJET-II-04 and SIYBLL2.3 are dominated by nondiffraction.

The identification of diffraction based on the rapidity gap technique has been investigated. We studied the 
performance of an effective selection criterion for diffractive events (central-veto selection): ``There are no charged particles ($N_{track}=0$) in the kinematic range $|\eta| <2.5$ and $p_T>100$ MeV." 
Such a selection has $\approx$100\% purity, independent of particle type, energy, and interaction model whereas selection efficiency increases from $\sim$30\% to 70\% with increasing energy. 
The surviving events from central-veto selection are mostly low-mass diffraction events in the phase space of $\log_{10}(\xi_{x}) < -5.5$. 
This indicates that the combined experiment can purify the detection of low-mass diffraction.     

Clearly, nondiffraction and diffraction have different contributions in the VF regions, while hadronic interaction models also exhibit big differences among each other. 
The rapidity gap measurement (central-veto technique) using central information is an effective way to identify diffractive events and classify the forward productions to nondiffraction and diffraction. 
Furthermore, combining the VF detector with central information offers a unique opportunity to both constrain the differential cross sections of low-mass diffraction events and to help to identify the inherent problems in the models corresponding to low-mass diffraction.   

\begin{acknowledgements}
We thank Sergey Ostapchenko and Tanguy Pieorg for illuminating discussions and suggestions. Part of this work was performed using the computer resources of the Institute for Cosmic Ray Research, University of Tokyo and the LHCf collaboration. We are grateful to C. Baus, T. Pierog, and R. Ulrich for providing the CRMC program codes. This work was supported by by Grant-in-Aids for Scientific Research by MEXT of Japan ( 23244050, 23340076, 23740183, 25287056, and 26247037 ). 
\end{acknowledgements}

%=========================================================================================
%=========================================================================================

\end{document}